\documentclass[aps,pre,amsmath,amssymb,nofootinbib,preprint]{revtex4-1}
\usepackage[utf8]{inputenc}
\usepackage{graphics}      
\usepackage{graphicx}      
\usepackage{longtable}     
\usepackage{url}           
\usepackage{indentfirst}
\newcommand{\beq}{\begin{equation}}
\newcommand{\eeq}{\end{equation}}
\newcommand{\bea}{\begin{eqnarray}}
\newcommand{\eea}{\end{eqnarray}}

\begin{document}
\title{Exact solutions of the Wheeler-DeWitt equation with ordering term in a dark energy scenario}

\date{\today}

\author{C. R. Muniz}
\email{celio.muniz@uece.br}
\affiliation{Grupo de F\'isica Te\'orica (GFT), Universidade
Estadual do Cear\'a, Faculdade de Educa\c c\~ao, Ci\^encias e Letras de Iguatu, Iguatu, Cear\'a, Brazil}
\author{H. R. Christiansen}
\email{hugo.christiansen@ifce.edu.br}
\affiliation{Grupo de F\'isica Te\'orica (GFT),  Instituto Federal de
Ciência, Educação e Tecnologia, CEP 62042-030, Ceará, Brazil}
\author{M. S. Cunha}
\email{marcony.cunha@uece.br}
\affiliation{Grupo de F\'isica Te\'orica (GFT), Centro de
Ci\^encias e Tecnologia, Universidade Estadual do Cear\'a, CEP 60714-903, Fortaleza, Cear\'a, Brazil}
\author{H. S. Vieira}
\email{horacio.santana.vieira@hotmail.com}
\affiliation{Instituto de Física de Sāo Carlos, Universidade de São Paulo, Caixa Postal 369, CEP 13560-970, São Carlos, São Paulo, Brazil}

\begin{abstract}
We investigate the quantum evolution of the universe in the presence of two types of dark energies. First, we consider the phantom class ($\omega<-1$)  which would be responsible for a super-accelerated cosmic expansion, and then we apply the procedure to an ordinary $\Lambda>0$ vacuum ($\omega=-1$). This is done by analytically solving the Wheeler-DeWitt equation with ordering term (WdW) in the cosmology of Friedmann-Robertson-Walker. In this paper, we find exact solutions in the scale factor $a$ and the ordering parameter $q$. For $q=1$ it is shown that the universe has a high probability of evolving from a big bang singularity. On the other hand, for $q = 0$ the solution indicates that an initial singularity is unlikely. Instead, the universe has maximal probability of starting with a finite well-defined size which we compute explicitly at primordial times. We also study the time evolution of the scale factor by means of the Hamilton-Jacobi equation and show that an ultimate {big rip} singularity emerges explicitly from our solutions. The phantom scenario thus predicts a dramatic end in which the universe would reach an infinite scale factor in a finite cosmological time as pointed by Caldwell et al. in a classical setup \cite{Caldwell}. Finally, we solve the WdW equation with ordinary constant dark energy and show that in this case the universe does not rip apart in a finite era.

\end{abstract}

\maketitle

\section{Introduction}

When Albert Einstein extended the field equations of General Relativity (GR) to include the cosmological constant (1917) he was seeking to dodge the collapse under the unavoidable attraction of gravity in what he believed to be a static universe. He could not foresee that this idea, considered by him the biggest blunder of his academic life \cite{Harvey}, would lead to a scenario of cosmic instability much more dramatic than the one he feared most. Although that decision made possible the desired time-independent solution, it described a state of unstable equilibrium irrespective of the existence of a critical mass density at which this repulsion could balance the gravitational attraction.  Since the cosmological constant acts like a repulsive force that increases with distance,  a slight expansion would rise the repulsion and decrease the attractive force making expansion accelerate.

Some years ahead, after the observation of the red-shift in the spectrum of galaxies located far outside our local group, Edwin Hubble announced (1929-1931) \cite{Hubble} that the universe was not shrinking but indeed uniformly expanding. But as above mentioned, the cosmological constant says more than this since it is responsible for an {accelerated} expansion. The confirmation of this phenomenon (1998) was at last determined from the Hubble diagrams of Type Ia supernovae \cite{SNIa,Perlmutter} and is one of the landmarks of astrophysics and cosmology considering the very proof of GR occurred 100 years ago in the luminous sky of Sobral (Ceará, Brazil).

The cosmological constant, $\Lambda > 0$ \cite{Sci312},  is associated with a type of energy that acts repulsively on a cosmic scale called {dark energy} for its  completely unknown nature.  The possibility that it could be derived from the quantum fluctuations of the vacuum associated with the fields existing in nature is still under discussion. This is the so-called cosmological constant problem or vacuum catastrophe. The name comes from the disagreement between the observed values of vacuum energy density (the small value of the cosmological constant) and the theoretical (large) value of zero-point energy suggested by quantum field theory \cite{SolaUnruh,Wang}.  Depending on the Planck energy cutoff and other factors, the discrepancy can be of more than 120 orders of magnitude, a state of affairs described as the largest gap between theory and experiment in all sciences \cite{CUP2006}. Notwithstanding these facts, there are novel proposals in order to solve this dramatic inconsistency by means of cancellations of the vacuum energy divergences considering supersymmetry \cite{Venturi} as well as more involved mechanisms  \cite{WU,Carlip}.

Dark energy is the form of energy that largely predominates ($68\% $) over the other types of energies in the present day observable universe (dark matter stays with $27\% $ and the rest is mostly hadronic matter, with just a bit of leptonic matter and electromagnetic radiation). The density of dark energy is very low ($7 \times 10^{-30} g/cm^3$) much less than the density of ordinary matter or dark matter within galaxies. However, it dominates the mass–energy of the universe because it seems to be everywhere in space. This would then explain the accelerated expansion of the universe.

The cosmological constant defines a frontier with other possible types of dark energy. There is a cosmological parameter that characterizes the types of energies that may exist in the universe in the simple form of a perfect fluid. This parameter appears in the state equation that relates the pressure of the fluid, $p$, and its density, $\rho$, in a relation of direct proportionality: $p=\omega \rho$. In most of the cosmological models the state parameter $\omega$ is a constant which, if negative, implies a negative pressure (energy density is supposed to be positive).
The standard accelerated expansion \cite{SNIa} is compatible with $-1<\omega<-1/3$. This range characterizes the so-called {\it quintessence} which, unlike the $\Lambda$ type dark energy would have variable energy density in space and time and could therefore  exert local effects on astronomical scales \cite{Celio}. It would however become less rarefied with the expansion of the cosmos as compared with ordinary matter and radiation. At the limit of $\omega=-1$, density does not vary with  expansion thus denoting the ordinary dark energy associated with $\Lambda$.

The possibility of a cosmic drama would start if dark energy happens to be associated with $\omega$ definitely {below} $-1$. In this case, the energy density would increase with expansion then causing more and more acceleration until the universe virtually reaches an infinite size in a finite cosmological time. This would have dramatic implications since the cosmological horizon would gradually decrease until collapsing to one point. In this process, not only would all galaxies, and even the stars of our Milky Way, progressively leave our field of observation (or any form of access), but would also be shattered due to the extraordinarily repulsive character of this form of energy sinisterly called {\it phantom} \cite{phantom} with the terrible final scenario here outlined and suggestively dubbed {\it big rip} by Caldwell \cite{Caldwell}.

Although the Planck satellite data are consistent with a $\Lambda$ type dark energy ($\omega=-1.03\pm 0.03$ \cite{Planck}), more recent observations of quasars at high redshifts, combined with those from the Cosmic Microwave Background and weak lensing, match with values of $\omega$ even below -1.5 \cite{Nature} indicating a rather high accelerated cosmic expansion.

In this paper, we speculate with an ideal theoretical setup consistent with pure dark energy in the form of a perfect fluid associated with a cosmological parameter $\omega <-1$ in order to test an extreme scenario, without worrying about the microscopic origin of such a fluid.
We start our analysis by considering the referred fluid with an arbitrary state parameter filling a flat universe, then specializing for $\omega=-3/2$ as in \cite{Caldwell}. It is noteworthy that this value is fully compatible with Risaliti and Lusso's recent findings \cite{Nature} and, significantly, allows the analytical treatment of the differential equations. Our set up has an important improvement over Caldwell's, namely the examination of some features of the quantum nature of the universe by means of the  Wheeler-DeWitt equation.

Regarding the WdW equation, we consider an additional term coming from the ambiguity in the ordering of the conjugated quantum momenta  related to the scale factor, namely $\hat{p}_a^2=-a^{-q}\partial/\partial a (a^q\partial/\partial a)$. Note that the quantum nature of the WdW equation holds even on a large scale \cite{Norbury} since there is no external environment through which the wave function of the universe could suffer decoherence to become classical like the objects present in our current world \cite{Yurov}. Here, we shall adopt a semi-classical approach by combining the study of the WdW equation with Hamilton-Jacobi's in order to find the time evolution of the scale factor of the universe which is absent in WdW.

We will show that the {big rip} scenario emerges out analytically from the  solutions of the WdW equation. Next, we will evaluate what happens in this scenario when instead of {phantom} energy we consider $\Lambda$ dark energy. In this case, we will show that the universe does not reach the {big rip} singularity.

The article is organized as follows. In section II, the WdW and Hamilton-Jacobi equations are studied in order to find the dynamics of the universe for dark energies associated with {phantom} and the cosmological constant respectively. In section III, we present the results and make the pertinent discussions. And finally, in section IV, we present our conclusions.

\section{A quantized universe: the Wheeler-DeWitt equation}

The equation of Wheeler and DeWitt (WdW) is about the wave function of the universe \cite{wdw,Wheeler, Misner}. It looks similar to the Schrödinger's equation in one dimension with the important difference that it does not take into account the time evolution. Considering the cosmology of Friedmann-Robertson-Walker (FRW), in the minisuperspace approximation the WdW equation can be written as
\begin{equation}
\left\{\frac{d^{2}}{da^{2}}+\frac{q}{a}\frac{d}{d a}-\frac{9\pi c^{4}a^{2}}{4\hbar^{2}G^{2}}\left[kc^{2}-\frac{8\pi Ga^{2}}{3c^{2}}(\sum_i\rho_{\omega_i})\right]\right\}\Psi(a)=0\ ,
\label{eq:WDE}
\end{equation}
where the wavefunction of the universe, $\Psi(a)$, is a function of the scale factor $a$ which is the relevant generalized coordinate. The parameter $k$ determines the curvature of the universe, and the different forms of energy can be written as
\begin{equation}
\rho_{\omega_i}=\frac{A_{\omega_i}}{a^{3(\omega_i+1)}}\ ,
\label{eq:WDE_density}
\end{equation}
where  $\omega_i$ are the state parameters which guarantee the proportionality between pressure and energy density for each form of energy taken as perfect fluids.
The second term, involving the (dimensionless) parameter $q$, comes from the ordering of the conjugated operators $\hat a$ and $\hat{p}_a$ where
$$\hat{p}_a^2=-a^{-q}\frac{\partial}{\partial a} (a^q\frac{\partial}{\partial a}).$$
It takes into account the ambiguity in the construction of the hamiltonian  \cite{ordering,Misnerb,Hawking}. Halliwell and Misner have shown that the ordering ambiguity in the WdW equation can be completely fixed by demanding invariance under field redefinition of both the three-metric and the lapse function for $D>1$, where $D$ is the number of gravitational coordinates in a mini-superspace model.
Following Ref. \cite{He}, we will work with a dynamical interpretation of the wavefuncion of the universe.

In this section, we will not take into account the energy density of the vacuum. This will be done later on.
We will now consider a flat \cite{Planck} $k=0$  pure dark energy scenario of the phantom class $\omega<-1$.
The WdW equation then reads
\begin{equation}
\frac{\partial^{2}\Psi(a)}{\partial a^{2}}+\frac{q}{a}\frac{\partial\Psi(a)}
{\partial a}+V_{eff}(a)\Psi(a)=0.
\label{WdW}
\end{equation}
The effective potential $V_{eff}(a)$ is initially given by
\begin{equation}
V_{eff}(a)=-A a^2 +B a^{4-3(\omega +1)},
\label{potential}
\end{equation}
where
\begin{equation}
A=\frac{9\pi c^{6}k}{4\hbar^{2}G^2},
\label{eq:A_FRW_universe}
\end{equation}
and
\begin{equation}
B=\frac{6\pi^{2}c^{2}}{\hbar^{2}G}A_{p},
\label{eq:B_FRW_universe}
\end{equation}
$A_p$ being a constant proportional to the {phantom} energy density at a given instant. The phantom hipotesis guarantees its own  predominance in a future cosmological time.

To find the solution of Eq. (\ref{WdW}), we first proceed to transform
$\Psi(a)=a^{(1-q)/2}\phi(a)$, which results in
\begin{equation}
a^2\phi''(a)+a\phi'(a)+\left[Ba^{3(1-\omega)}-\left(\frac{1-q}{2}\right)^2\right]\phi(a)=0.
\label{EDOTransf1}
\end{equation}
Now, defining
$z=2\left[\sqrt{B}/3(1-\omega)\right]a^{3(1-\omega)/2}$ we obtain
\begin{equation}
z^2\phi''(z)+z\phi'(z)+\left[z^2-\left(\frac{1-q}{3(1-\omega)}\right)^2\right]\phi(z)=0.
\label{EDOTransf2}
\end{equation}
This is a Bessel differential equation
whose general solution is given in terms of the Bessel functions of first and second kind, $J_{\nu}(z)$ and $Y_{\nu}(z)$ where $\nu=\frac{1-q}{3(\omega-1)}$. Note that we can put the second solution in terms of $J_{-\nu}(z)$ which will be useful as we shall see next.

Returning to the original variables the solution to (\ref{EDOTransf1}) reads
\begin{eqnarray}
\Psi(a)&=&C_1(\omega,q) B^{\frac{1-q}{6(1-\omega)}} a^{\frac{1-q}{2}}
J_{\frac{1-q}{3(\omega-1)}}\left[\frac{2\sqrt{B} a^{3/2(1-\omega)}}{3(1-\omega)}\right]\nonumber\\
&+&C_2(\omega,q) B^{\frac{1-q}{6(1-\omega)}}  a^{\frac{1-q}{2}}
J_{\frac{q-1}{3(\omega-1)}}\left[\frac{2\sqrt{B} a^{3/2(1-\omega)}}{3(1-\omega)}\right],
\label{SolucaoGeral}
\end{eqnarray}
where $C_{1}$ and $C_2$ are normalization constants.





For $q=0$ the second solution is multivalued at the origin for the index of $J_{\nu}(z)$ becomes a negative non-integer number. Thus, this choice picks the first solution which is also free of divergences at infinity. We also note that $q=0$ admits the Hartle-Hawking contour condition,  compatible with the universe's wave function square growing from the origin \cite{Hawking2}. There is no value of $q$ picking only the second solution deprived of complications with divergences.
The only value of $q$ for which the wave function is finite at the origin is $q=1$, corresponding to the Linde's boundary condition (decreasing mode from the origin) \cite{Vilenkin}. In fact, for this value both solutions are well behaved with no boundary divergences.

In order to exactly normalize the solutions, we set $\omega=-3/2$ \cite{Nature}. In the first case, $q=0$, after integration from $a=0$ to $a=\infty$, the normalized wave function of the universe results
\begin{equation}\label{UniverseWaveFunction0}
\Psi(a)=\sqrt[4]{\pi} 15^{\frac{7}{30}}2^{\frac{1}{30}}\sqrt[15]{B}\sqrt{\frac{\Gamma(\frac{11}{15})\Gamma(\frac{13}{15})}{\Gamma(\frac{7}{30})\Gamma(\frac{2}{5})}}J_{2/15}\left[\frac{4\sqrt{B} a^{\frac{15}{4}}}{15}\right].
\end{equation}
For $q=1$ the wave function of the universe reads exactly
\begin{equation}\label{UniverseWaveFunction}
\Psi(a)=\frac{\sqrt[4]{\pi} 15^{\frac{11}{30}}\Gamma(\frac{13}{15})}{2^{\frac{7}{30}}\sqrt{\Gamma(\frac{2}{15})\Gamma(\frac{11}{30})}}\sqrt[15]{B}J_0\left[\frac{4\sqrt{B} a^{\frac{15}{4}}}{15}\right].
\end{equation}

\begin{figure}[!h]
	\centering
	\includegraphics[scale=0.40]{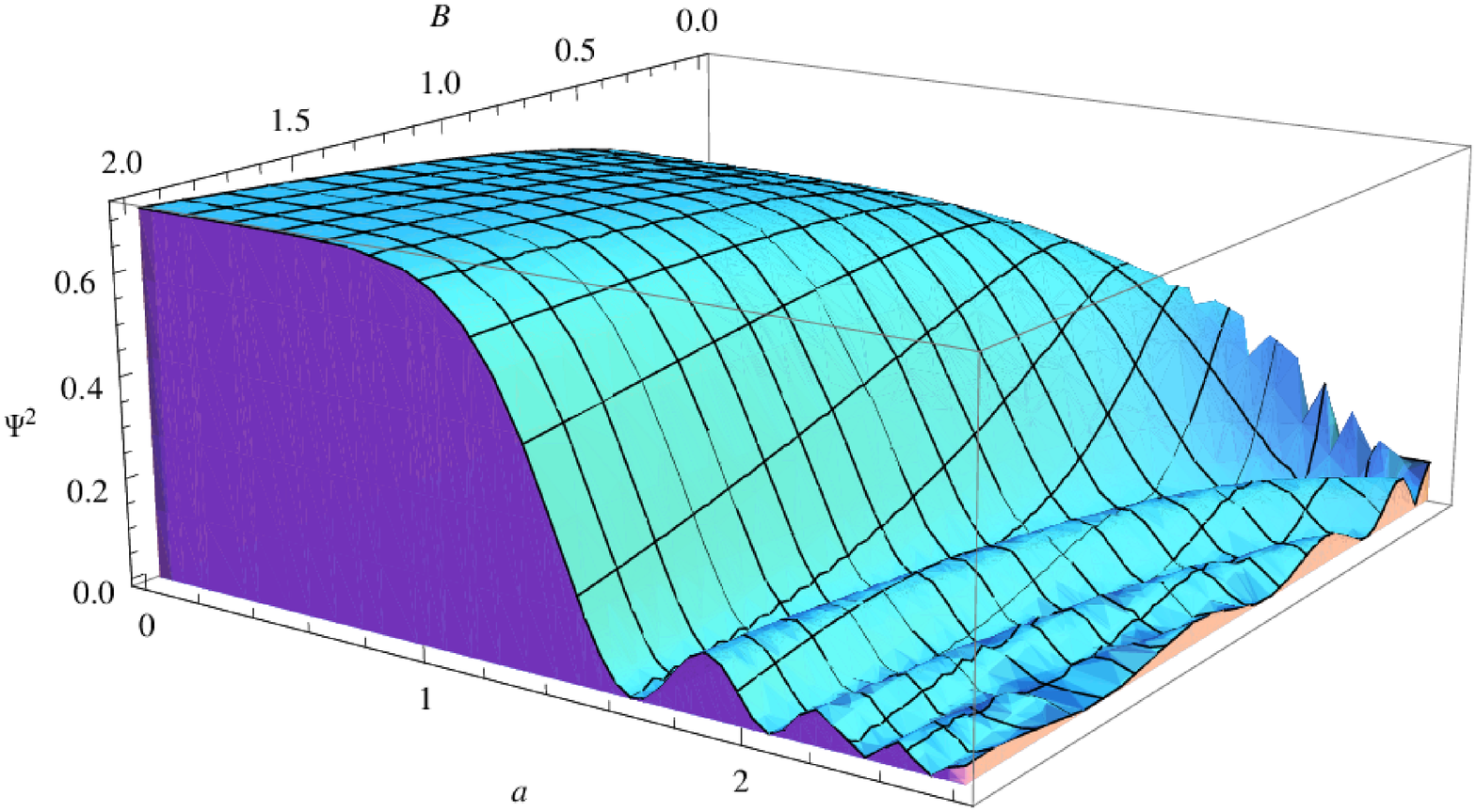}
	\caption{Square modulus, $|\Psi|^2$, as a function of $a$ and $B$, for $q=1$.}
	\label{FIG3D1}
\end{figure}

\begin{figure}[!h]
	\centering
	\includegraphics[scale=0.40]{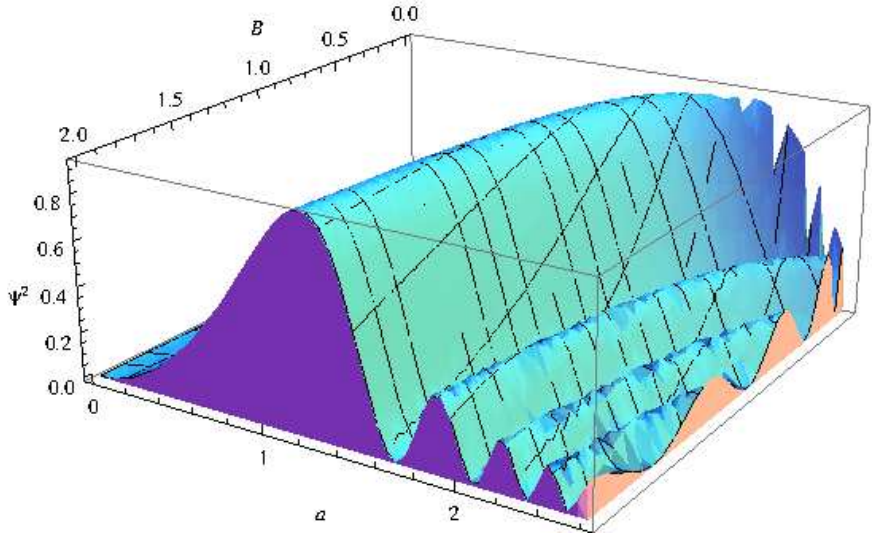}
	\caption{Square modulus, $|\Psi|^2$, as a function of $a$ and $B$, for $q=0$.}
	\label{FIG3D0}
\end{figure}

In Figs. (\ref{FIG3D1}) and  (\ref{FIG3D0}), we plot the square modulus of $\Psi$
as a function of $a$ and $B$.

According to Fig. \ref{FIG3D1}, for $q=1$ the universe starts with a high probability of evolving from a big bang singularity, where $a=0$, and such a probability grows with the amount of phantom energy.

On the other hand, it is interesting to observe that in Fig. \ref{FIG3D0}, stemming from Eq.(\ref{UniverseWaveFunction0}), the universe presents a very high probability of coming into existence with a well defined size  whose probability is higher the lower the phantom energy content is, and therefore an initial singularity is unlikely.  In this case, we can obtain a relationship between the phantom energy density of the universe in a specific instant, $A_p$, and the initial value for the scale factor, $a_0$, by calculating the maximal probability at this value from $\partial_a |\Psi(a)|^2=0$.
Now, for a high phantom energy density $A_p$ we obtain, near the origin,
\begin{equation}
a_0=\left(\frac{85 \hbar^{2}G \left[2\,\Gamma\left(\frac{2}{15}\right)+15\, \Gamma\left(\frac{17}{15}\right)\right] \Gamma\left(\frac{32}{15}\right)}{4 \pi^{2}c^{2}A_{p} \{285\, \Gamma\left(\frac{17}{15}\right) \Gamma \left(\frac{32}{15}\right)+\Gamma\left(\frac{2}{15}\right) \left[34\, \Gamma\left(\frac{17}{15}\right)+8\, \Gamma\left(\frac{32}{15}\right)\right]\}}\right)^{2/15}.
\label{InitialRadius}
\end{equation}
The above expression shows that the bigger is the density of phantom energy, the smaller is the size of the starting universe. Notice that in a purely classical scenario ($\hbar\to 0$) it results $a_0\to 0$ and we thus obtain the initial singularity back.
Nevertheless, provided $|\Psi(a)|^2$ represents the probability density of the universe staying in the state $a$ during its evolution \cite{He}, the resulting wave function presents no big bang singularity but reveals a big rip divergence just as the classical model \cite{Caldwell}. In what follows we will further argue about this point.

\section{The dynamics of the universe}

Although the WdW equation does not explicitely involve time evolution, it has a dynamical interpretation which has been recently discussed in detail  \cite{He}. Thus, in order to get some insight  into the time evolution of the universe we will follow the method described in \cite{Horacio}, which consists in a semiclassical Hamilton-Jacobi's equation approach. As the wavefunction $\Psi(a)$ depends on just the scale factor of the universe, it can be rewritten as
\begin{equation}
\Psi(a)=R(a)\ \mbox{e}^{iS(a)}\ ,
\label{eq:psi_expansion}
\end{equation}
where $R$ and $S$ are real functions.

From quantum mechanics, the probability current is given by
\begin{equation}
j^{a}=\frac{i\hbar}{2\mu}[\Psi^{*}(\partial_{a}\Psi)-\Psi(\partial_{a}\Psi^{*})]\ ,
\label{eq:current_density_FRW_universe}
\end{equation}
guaranteeing the conservation law
\begin{equation}
\partial_{a}j^{a}=0\ .
\label{eq:conserved_FRW_universe}
\end{equation}
By replacing Eq.~(\ref{eq:psi_expansion}) in Eq.~(\ref{eq:current_density_FRW_universe}), we obtain
\begin{equation}
j^{a}=-\hbar R^{2}\frac{\partial S}{\partial a}\ ,
\label{eq:j_1_FRW_universe}
\end{equation}
and integrating Eq.~(\ref{eq:conserved_FRW_universe}), we get
$
j^{a}=C_{0}\ ,
\label{eq:j_2_FRW_universe}
$
where $C_{0}$ is an arbitrary constant. Thus,
\begin{equation}
-\hbar R^{2}\frac{\partial S}{\partial a}=C_{0}\ .
\label{eq:result_1_FRW_universe}
\end{equation}

Now, we can use the Hamilton-Jacobi formalism of quantum mechanics to write the following relation between action and canonical momentum
\begin{equation}
p_{a}=\frac{\partial S}{\partial a}=\frac{\partial L}{\partial \dot{a}}=-\frac{3 \pi c^{2}}{2G}\dot{a}a\ ,
\label{eq:action_momentum_FRW_universe}
\end{equation}
where $L$ is the Lagrangian of the Friedmann-Robertson-Walker universe.
Thus, from Eqs.~(\ref{eq:result_1_FRW_universe}) and (\ref{eq:action_momentum_FRW_universe}), we come to
\begin{equation}
R^{2}=|\Psi|^2=C_{0}\frac{2 G}{3 \hbar \pi c^{2} \dot{a} a}\ .
\label{eq:result_2_FRW_universe}
\end{equation}

\subsection{The big rip}

We can now obtain the evolution of the scale factor of the universe over cosmological time. Substituting the solution (\ref{UniverseWaveFunction}) in Eq.  (\ref{eq:result_2_FRW_universe}), and integrating from $t=0$ to a certain $t_{rip}$
and from $a=0$ to $a=\infty$, we obtain
\begin{equation}
t_{rip}=\frac{15^{19/15}}{5\times2^{23/15}}\frac{\Gamma(13/15)^2\Gamma(7/30)\Gamma(4/15)}{\Gamma(11/15)^2\Gamma(2/15)\Gamma(11/30)}\frac{\pi c^2\hbar}{C_0 G B^{4/15}},
\label{FiniteTime}
\end{equation}
with $B$ given by Eq.(\ref{eq:B_FRW_universe}).
	
On the other hand, assuming $q=0$, which selects from Eq. (\ref{SolucaoGeral}) the wave function of the universe with high probability of starting with a finite radius, we replace this solution in Eq. (\ref{eq:result_2_FRW_universe}) and integrate from $a=a_0$ to $a=\infty$. We find once more a finite value for $t_{rip}$ but in a more involved  expression on $B$ and $A_p$. However,  as a function of $a_0$, $t_{rip}$ is quite simple, increasing (decreasing) monotonically with the minimal radius (actual phantom density).  Both solutions for the cosmological time as a function of the scale factor result therefore in an universe expanding infinitely in a finite era. This truly strange and disturbing scenario predicts a dramatic fate for the universe.

As we have seen, although the big rip is an ultra-high scale phenomenom, the time to reach such singularity depends on the Planck constant $\hbar$. Such result thus reinforces the argument that even on very large scales the universe is essentially of a quantum nature.

\subsection{Cosmological constant: a happier cosmic destiny}

We will now analyze the situation in which the energy density of the universe is associated with the ordinary vacuum energy density, $\rho_v$, in the form of a positive cosmological constant, $\Lambda$. In this case, the energy density remains the same while the universe expands, and the actual potential will be $V_{eff}(a)=-B_{\Lambda}a^4$, where $B_{\Lambda}=\frac{3\pi c^6}{4\hbar^2G^2}\Lambda$.

The solution of Eq. (\ref{WdW}) can be once again analytically given in terms of Bessel functions
\begin{eqnarray}
\Psi(a)&=&C_1\ B_{\Lambda}^{\frac{1}{2} \left(\frac{q}{6}-\frac{1}{6}\right)+\frac{1-q}{6}}
a^{3 \left(\frac{q}{6}-\frac{1}{6}\right)-q+1} \Gamma \left(\frac{7}{6}-\frac{q}{6}\right)
J_{\frac{1-q}{6}}\left(\frac{\sqrt{B_{\Lambda}} a^3}{3}\right)\nonumber \\
&+&C_2\ B_{\Lambda}^{\frac{1}{2} \left(\frac{1}{6}-\frac{q}{6}\right)} a^{3 \left(\frac{1}{6}-\frac{q}{6}\right)}
\Gamma \left(\frac{q}{6}+\frac{5}{6}\right) J_{\frac{q-1}{6}}\left(\frac{\sqrt{B_{\Lambda}} a^3}{3}\right).
\end{eqnarray}

Taking the value $q=0$ and the first of the L.I. solutions above for the reasons already explained, we can make the substitution in the Eq. (\ref{eq:result_2_FRW_universe}), integrating  once again  in the range $a\in[0,\infty)$, and $t\in[t_0,t]$. We found as expected that $t-t_0$ diverges, meaning that the universe tends to an infinite size in infinite time. We obtain the same behaviour for $q=1$.

\section{Results and Discussion}

\begin{figure}[!h]
\centering
\includegraphics[scale=0.65]{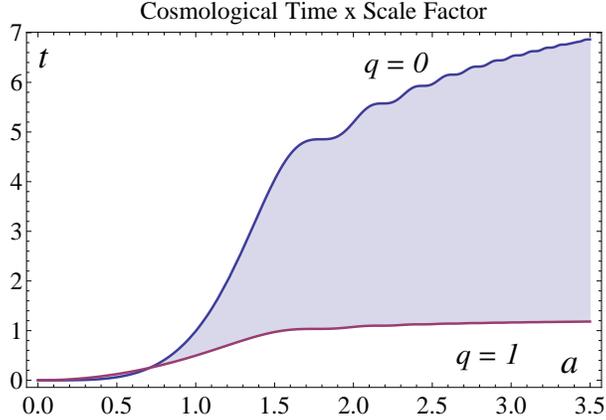}
\caption{Cosmological time, $t$, as a function of the scale factor, $a$, for $q=0$ and $q=1$.
All the constants are set to one and $t_0=0$.}
 \label{FIG1}
\end{figure}

\begin{figure}[!h]
\centering
\includegraphics[scale=0.7]{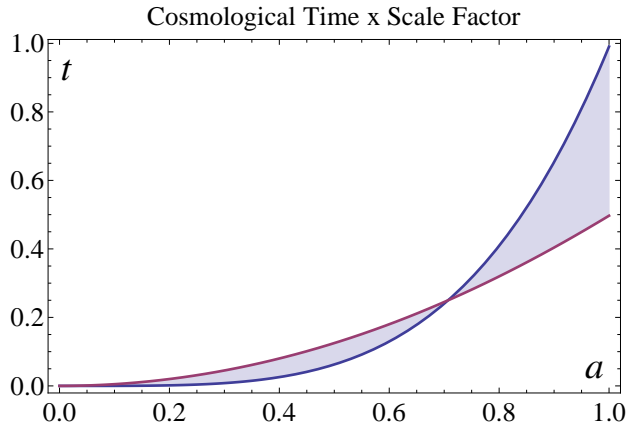}
\caption{Detail of the above plot at primordial times. }
 \label{FIG2}
\end{figure}

In Figs. \ref{FIG1} and \ref{FIG2} we plotted the cosmological time as a function of the scale factor,   obtained from integration of Eq. (\ref{eq:result_2_FRW_universe}) for arbitrary $a$ and $t$, illustrating what happens with the expansion of a universe mastered by phantom energy. The $q=0$ solution corresponds to a universe starting its existence with a very pronounced expansion, even more than in the case  $q=1$. After some period, the expansion still grows hard but it is relativelly attenuated and starts an era of ripple.

The rapid expansion recorded up to the first shoulder of the curve corresponds to the beginning of times. Although tempting,  we cannot directly associate it to the usual cosmic inflation because we did not include any other ingredient such as the {\it inflaton} and other forms of energy certainty present in the primordial universe.

\begin{figure}[!h]
\centering
\includegraphics[scale=0.3]{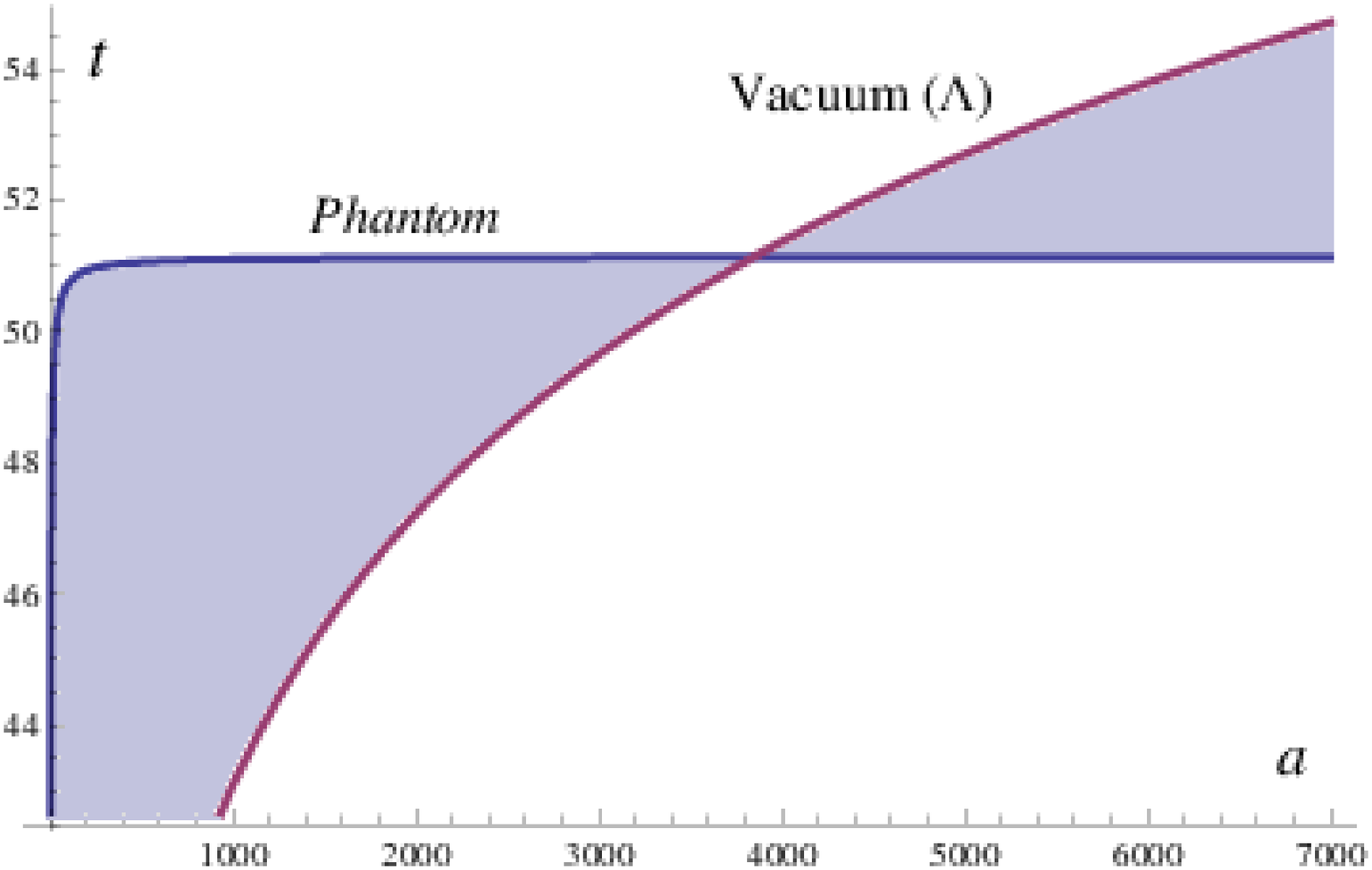}
\caption{Comparison of the cosmological time solutions $t(a)$ between universes with predominance of phantom energy (blue) and a vacuum cosmological constant (red). All the constants are set to one and $t_0=0$.}
 \label{FIG3}
\end{figure}
According to Fig. (\ref{FIG1}), at a certain time of the cosmic history, when the phantom energy content predominates over the others, there is a succession of accelerations and decelerations in the expansion of the universe, very visible in the $q=0$ case. The amplitude of these oscillations decreases as the universe expands faster and faster, until it tends to an infinite size in a finite time of existence, see Eq.(\ref{FiniteTime}). It is interesting to compare these scale factor oscillations with those found from supernova type Ia data \cite{Mead}. In Fig.(\ref{FIG2}) we see the detail of the Fig. (\ref{FIG1}) at primordial times.

From Eq.(\ref{FiniteTime}), we can see that the higher the phantom energy content is the shorter the time to reach the big rip since nothing will resist this unbridled expansion, not even hadrons as the very fabric of space-time is about to break. Furthermore, Fig. \ref{FIG1} still reveals that the case $q=1$ is even more dramatic, because the  big rip is reached even faster for the same initial phantom energy content $A_p$.
We suggest looking at Table 1 in \cite{Caldwell} where an interesting cosmological-time scenario is shown for a phantom energy set up with $\omega=-3/2$. We have specialized to this value in order to compare thta results with our analytical calculations in a quantum cosmological approach.

\section{Concluding remarks}

We have found the solutions of the Wheeler-DeWitt equation describing the quantum mechanical wavefunction of the universe in two dark energy scenarios: the usual cosmological constant background and the phantom. We have worked in a FRW cosmology with a term characterizing the ambiguity in the ordering of the conjugate operators associated with the scale factor. Such term depends on a parameter $q$ which makes possible avoiding boundary divergences in $\Psi(a)$ for $q=0$ and $q=1$.

Although we have not included classical matter or radiation, which would be necessary in a reliable universe at early stages, in the far evolved regime phantom energy would be predominant. Under this assumption we were able to obtain the exact solutions of the WdW equation for arbitrary state parameters. Actually, our focus was the phantom energy density range given by $\omega < -1.3$ in a flat space, as suggested from the analysis of the last experimental data in \cite{Nature}.

By specializing for $\omega=-3/2$ \cite{Caldwell} we analytically showed that for $q = 0$ the universe comes into existence with a finite nonsingular size. This prospect gets more probable the higher the phantom energy content is, and a big bang birth with $a = 0$ results unlikely (see Fig. \ref{FIG3D0}).
Indeed, we have calculated the value of the initial scale factor of the universe as a function of the phantom energy density and the Planck constant, see Eq.(\ref{InitialRadius}). For $q=1$, on the other hand, the universe has a high probability of evolving from a big bang singularity, where $a=0$, and such a probability grows with the amount of phantom energy, see Fig. \ref{FIG3D1}.
Regarding large scales, through a semiclassical approach via Hamilton-Jacobi dynamics we have shown that the scale factor tends to infinity in a finite cosmological time for both values of $q$. This confirms what purely classical models based on the solutions of Friedmann's equation have already predicted.

We also showed that even in the large scale the universe is apparently quantum in nature since the time to reach the rip era also depends on $\hbar$, see Eq.(\ref{FiniteTime}). Such a unique eventual state is the culmination of cosmic history in which a particular form of dark energy (the phantom) dominates over all the others. At the end of this cosmological era all the material structures, no matter their degree of cohesion, from clusters of galaxies to subatomic particles, will be shattered due to the extremely repulsive character that this energy assumes (as the universe expands the phantom energy becomes more dense and more destructive).

We also examined the case where the dark energy is in the form of the Einstein's cosmological constant $\Lambda>0$, which corresponds exactly to $\omega=-1$. In this case we obtain a singular beginning, as expected, and the {big rip} finale is excluded because the scale factor tends to infinity in an infinite cosmological time (see Fig. \ref{FIG3}).

It is noteworthy that we have been able to find time oscillations in the scale factor of the universe whose amplitude decreases with the cosmological time, see Fig. \ref{FIG1}. This matches the recent findings of Ringermacher and Mead \cite{Mead}.

\section*{Acknowledgements} The authors thank CNPq and FUNCAP for their partial support under the grant PRONEM PNE-0112-00085.01.00/16. MSC is also under the grants 433168/2016-1 and 314183/2018-3. HSV is funded by the Coordenação de Aperfeiçoamento de Pessoal de Nível Superior (CAPES) - Finance Code 001.


\begin{thebibliography}{10}
\section*{References}
\bibitem{Caldwell} Caldwell, R. R., Kamionkowski, M., and Weinberg, N. N. ``Phantom Energy: Dark Energy with $\omega<-1$ Causes a Cosmic Doomsday'', Phys. Rev. Lett. {\bf91}, 071301 (2003).
\bibitem{Harvey} Harvey, A.,  Schucking, E. ``Einstein'\,s mistake and the cosmological constant'', Am. J. Phys. {\bf 68}, 8, 723 (2000).
\bibitem{Hubble} Hubble, E. "A relation between distance and radial velocity among extra-galactic nebulae", PNAS 15, 3, 168 (1929).
\bibitem{SNIa} Riess, A. G.  \textit{et al.}  "Observational Evidence from Supernovae for an Accelerating universe and a Cosmological Constant". Astrophys. J. 116 (3) 1009-1038 (1998).
\bibitem{Perlmutter} S. Perlmutter, \textbf{et al.} ''Measurements of Omega and Lambda from 42 high redshift supernovae''. Astrophys. J. 517, 565 (1999).
\bibitem{Sci312} Steinhardt, P. J., Turok, N. "Why the cosmological constant is small and positive". Science 312, 5777, 1180-1183 (2006).
\bibitem{SolaUnruh} Sola, J. "Cosmological constant and vacuum energy: Old and new ideas", J. Phys.: Conf. Ser. 453, 012015 (2013);
\bibitem{Wang} Wang, Q., Zhu, Z., and Unruh, W. G. "How the huge energy of quantum vacuum gravitates to drive the slow accelerating expansion of the universe", Phys. Rev. D 95, 103504 (2017).
\bibitem{CUP2006} Hobson, M. P, Efstathiou, G. P., Lasenby, A. N. General Relativity: An introduction for physicists (Reprint ed.). Cambridge University Press (2006). p.187.
\bibitem{Venturi} A.Yu. Kamenshchik, A.A. Starobinsky, A. Tronconi, T. Vardanyan, G. Venturi, ``Pauli-Zeldovich cancellation of the vacuum energy divergences, auxiliary fields and supersymmetry'', Eur.Phys. J. C \,{\bf 78}, 200 (2018).
\bibitem{WU} Q. Wang, W.G. Unruh, ''Vacuum fluctuation, micro-cyclic universes and the cosmological constant problem''. arXiv:1904.08599v2;
\bibitem{Carlip} S. Carlip, ''Hiding the Cosmological Constant'', Phys. Rev. Lett. 123, 131302 (2019).
\bibitem{Celio} V.B. Bezerra, H.R. Christiansen, M.S. Cunha, C.R. Muniz, M.O. Tahim, ''Thermal Casimir effect in Kerr spacetime with quintessence and massive gravitons'', Eur. Phys. J. {\bf C77} (2017) 787;
Bezerra, V.B., Cunha, M.S., Freitas, L.F.F.,  Muniz, C.R., Tahim, M.O.``Casimir effect in the Kerr spacetime with quintessence'', Mod. Phys. Lett. {\bf A32}, 01, 1750005 (2017).
\bibitem{phantom}  Caldwell, R. R. ''A phantom menace? Cosmological consequences of a dark energy component with super-negative equation of state'' Phys. Lett. B 545 (2002) 23-29.
\bibitem{Planck} Aghanin, N. {\it et al.}, Planck Collaboration, arXiv:1807.06209v1.
\bibitem{Nature} Risaliti, G.  and Lusso, E.``Cosmological constraints from the Hubble diagram of quasars at high redshifts'', Nature Astr. {\bf 3}, 272 (2019).
\bibitem{Norbury} Norbury, J. W. ``From Newton's laws to the Wheeler-DeWitt equation'', Eur. J. Phys. {\bf19}, 143 (1998).
\bibitem{Yurov} Yurov, A. and Yurov, V. ``The Day the universes Interacted: Quantum Cosmology without a Wave function''.
\bibitem{wdw} DeWitt, B. S. Phys. Rev. 160, 1113 (1967).
\bibitem{Wheeler} Wheeler, J. A. \textit{In}: Battelle Rencontres: 1967 Lectures in Mathematics and Physics, edited by DeWitt, C.  and  Wheeler, J. A. (Benjamin, New York, 1968).
\bibitem{Misner} Misner, C. W. \textit{In}: Magic Without Magic: John Archibald Wheeler, edited by Klauder, J. R. (Freeman, San Francisco, 1972).
\bibitem{ordering} Halliwell, J. J. Phys. Rev. D 38, 2468 (1988).
\bibitem{Misnerb} Misner, C. W. \textit{In}: Relativity, edited by Carmeli, M., Fickler, S. I., Witten, L. (Plenum,New York, 1970);
\bibitem{Hawking} Hawking, S. W., Page, D. N. Nucl. Phys. B 264, 185 (1986).
\bibitem{He} He, D., Gao, D., and Cai, Q., ``Dynamical interpretation of the wavefunction of the universe'', Phys. Lett.{\bf B748}, 361 (2015).
\bibitem{Kung} Kung, J. H. Gen. Rel. Grav. \textbf{27} (1995) 35.
\bibitem{Hawking2} J. B. Hartle and S. W. Hawking, Phys. Rev. {\bf D28}, 2960 (1983).
\bibitem{Vilenkin} Vilenkin, A. ``The quantum cosmology debate'', AIP Conference Proceedings {\bf478}, 23 (1999).
\bibitem{Barvinsky} Barvinsky, A. O., ``Unitary approach to quantum cosmology'', Phys.Rept., {\bf 230}, 237 (1993).
\bibitem{Horacio} Vieira, H. S.  and Bezerra, V. B.``Class of solutions of the Wheeler-DeWitt equation in the Friedmann-Robertson-Walker universe'', Phys. Rev. D \textbf{94},  023511 (2016).
\bibitem{Mead} Ringermacher, H. I. and Mead, L. R. ``Observation of Discrete Oscillations in a Model-Independent Plot of Cosmological Scale Factor Versus Lookback time and Scalar Field Model'', Astron. J. {\bf149}, 137 (2015).

\end{thebibliography}
\end{document}